\documentclass[12pt,preprint]{aastex}

\slugcomment{Submitted to Astrophysical Journal}

\shorttitle{C/O ratios in M dwarfs}
\shortauthors{Nakajima \& Sorahana}

\begin{document}


\title{Carbon-to-Oxygen Ratios in M dwarfs and Solar-type Stars}


\author{Tadashi Nakajima\altaffilmark{1}}
\affil{Astrobiology Center, 2-21-1, Osawa, Mitaka, Tokyo, 181-8588, Japan}
\email{tadashi.nakajima@nao.ac.jp}


\and

\author{Satoko Sorahana}
\affil{Department of Astronomy, Graduate School of Science, The University of Tokyo, 
7-3-1 Hongo, Bunkyo-ku, Tokyo, 113-0033, Japan}
\email{sorahana@astron.s.u-tokyo.ac.jp}


\altaffiltext{1}{National Astronomical Observatory of Japan, 2-21-1, Osawa, Mitaka, Tokyo, 181-8588, Japan}


\begin{abstract}

It has been suggested that high C/O ratios ($>0.8$) in circumstellar disks lead to the formation of
carbon dominated planets. Based on the expectation that elemental abundances in the stellar photospheres
give the initial abundances in the circumstellar disks, the frequency distributions of C/O ratios of solar-type
stars have been obtained by several groups. The results of these investigations are mixed. Some 
find C/O $>$0.8 in more than 20\% of stars, and C/O $>$1.0 in more than 6\%. Others find C/O $>0.8$
in non of the sample stars. These works on solar-type stars are all differential abundance analysis
with respect to the Sun and depend on the adopted C/O ratio in the Sun.  
Recently a method of molecular 
line spectroscopy of M dwarfs with which carbon and oxygen abundances are derived respectively
from CO and H$_2$O lines in the $K$ band, has been developed. 
{ The resolution of the $K$ band spectra is 20,000.}
Carbon and oxygen abundances of
46 M dwarfs have been obtained by this non-differential abundance analysis.
Carbon-to-oxygen ratios in M dwarfs derived by this method are more robust than those in solar-type
stars derived from neutral carbon and oxygen lines in the visible spectra due to the difficulty
in the treatment of oxygen lines.
We have compared the frequency distribution of C/O distributions in M dwarfs with those of
solar-type stars and have found that low frequency of high C/O ratios is preferred.

\end{abstract}





\section{Introduction} 


The study of elemental abundances in the stellar atmosphere is an
important part of modern astronomy. Carbon and oxygen
are among the most important elements and the carbon-to-oxygen ratio 
is expected to be the essential parameter of the atmosphere of the star hosting
planets, which controls the nature of the planets.
In the early 2000's, possible existence of carbon dominated planets with
carbon dominated silicates
was discussed 
\citep{gai00,kuc05}.

In the early 2010's, 
some determinations of the carbon-to-oxygen ratios in solar-type stars
found that $\sim$25\% of nearby stars have C/O $>$ 0.8 
and $\sim$8\% have C/O $>$ 1.0 
\citep{bon10,del10,pet11}. 
\citet{bon10} suggested that the mineralogy of planets formed in the environmental gas
with C/O $>0.8$ will make carbon dominated rocky planets, rather than oxygen dominated planets.
These large fractions of high C/O ratio stars were criticized
by \citet{for12} who pointed out some issues. 
{ First the frequency of high C/O is inconsistent with relatively small fraction of dwarf carbon stars
($<10^{-3}$) in large samples of low-mass stars}. Second, these high C/O ratios are
overestimated. { The possible reasons for this overestimation are a high C/O ratio for
the Sun used for differential abundance analysis { and} the treatment of
a Ni blend that affects the O abundance. }

A careful analysis of carbon and oxygen abundances of FG main sequence stars in the solar neighborhood
was presented by \citet{nis14} by taking into account the non-LTE effects in the model atmosphere
and a critical analysis of the Ni blend to the forbidden [OI] line at 6300\AA. 
Among 66 nearby disk stars for which
$\lambda$7774 OI triplet was analyzed with the non-LTE analysis, no star with C/O$>0.8$ was found.   
{ \citet{nis14} conclude that C/O does not exceed 0.8, and this result seems to exclude formation of 
carbon dominated planets.}

Star formation history of M dwarfs is expected not to be so different from that of solar-type stars
and they can also be used as probes of high C/O in the solar neighborhood.
The spectra of M dwarfs include absorption bands of O-containing molecules such as TiO, VO,
and CaOH, which are sensitive to the available O abundance, and hence C/O.
In a two dimensional space span by TiO and CaH indices, \citet{gai15} and \citet{giz16} plot
a sequence of M dwarfs from which a small number of stars deviate in the direction of
higher C/O.  \citet{gai15} estimates that high C/O$\sim 1$ stars constitute less than 
$6 \times 10^{-4}$ of M dwarfs with 99\% confidence, while \citet{giz16} show that 
a high C/O ratio (0.9) is less than 1\%.

Recently carbon and oxygen abundances of 46 M dwarfs have been obtained respectively from CO
and H$_2$O molecular lines { derived from $K$ band spectra obtained at the Subaru telescope with
a resolution of 20,000 } \citep{tsu14,tsu15,tsu16}. 
In this paper,
we present the C/O ratios of these M dwarfs and examine
the kinematics of the sample in terms of stellar population 
in \S2 and compare the distribution of C/O of these M dwarfs
with those of solar-type stars obtained by others using the Kolmogorov-Smirnov test in \S3.
Implications of the comparison results are discussed in \S4.









\section{Abundances and kinematics of the M dwarf sample}


{ While abundance indicators of carbon and oxygen lines in solar-type stars are sensitive
to physical conditions of atmospheric models, there are numerous lines of stable CO and H$_2$O
in the $K$ band as abundance indicators in M dwarfs, which are insensitive to detailed physical
conditions of atmospheric models as shown graphically in Figure 13 of \citet{tsu15}.}
However, spectroscopic analysis of M dwarfs in the $K$ band has been deemed difficult 
{ due to the severe veiling of the spectra by H$_2$O bands, which hides the
true continuum necessary to measure equivalent widths.} 
\citet{tsu14} and \citet{tsu15} have overcome this problem in the following manner.
They first examine the nature of veil opacities and then modify the way to measure equivalent widths.
{ The progress in molecular line database such as HITEMP \citep{rot10}
has reached the point that it allows the evaluation of the pseudo-continuum level of M dwarf spectra
fairly accurately. They reformulate the spectroscopic analysis so as to refer to the pseudo-continuum
defined by the molecular veil opacities, in stead of referring to the true continumm}


First, the determination of carbon abundance from
the ro-vibrational lines of the CO(2-1) band is { described.}
{ Since the continuum of the observed M dwarf spectrum is depressed by numerous weak lines of
H$_2$O, what can be seen is only  the pseudo-continuum. In the case of the
model spectrum of the M dwarf, not only the true continuum, but also the 
pseudo-continuum can easily be
evaluated owing to the recently available H$_2$O line database.
Then the analysis of the M dwarf spectrum can be performed by referring to the pseudo-continuum levels
both on the observed and theoretical spectra. 
Therefore the abundant CO lines are superb carbon abundance indicators, since most of the carbon
atoms are in CO molecules, which are stable at low temperatures.} 
Next, the determination of oxygen abundance from the ro-vibrational lines
of H$_2$O is described.
The procedure is similar to the case of CO lines. { H$_2$O lines in the M dwarf spectrum 
are heavily blended with each other and the true continuum level is hidden by the collective absorption
of H$_2$O lines themselves.} However, it is possible to perform the analysis of H$_2$O lines
by referring to the pseudo-continuum consistently defined on the observed and theoretical spectra.
In the atmosphere of cool M dwarfs, oxygen atoms are first used to form CO molecules. Then almost
all the oxygen atoms left after the formation of CO molecules are used to form stable H$_2$O molecules,
which are unaffected by the uncertainties due to the imperfection in photospheric models.
{ Thus abundant H$_2$O lines are very good oxygen abundance indicators.}
{ It would be convenient if carbon and oxygen abundances for the Sun could be derived from  
a $K$ band spectrum as for M dwarfs, but it is impossible, since CO and H$_2$O molecules dissociate in the atmosphere of the Sun.}

The derived carbon and oxygen abundances which are compiled
from \citet{tsu14,tsu15,tsu16} for 46 M dwarfs are given in Table 1.
One of the important characteristics of this data set is that none of the M dwarfs
has a C/O ratio greater than 0.8.


The sample M dwarfs are nearby (d$ <$ 20 pc), and expected to be mostly thin disk stars
in terms of stellar population.
To confirm this expectation explicitly, we have analyzed their kinematics in the Galaxy in terms of
space velocities in the local standard of rest. 
{ The space velocities are calculated from proper motions, parallaxes, and radial velocities
obtained from SIMBAD using the transformation formula given by \citet{joh87}.}
These space velocities are given in Table 1 and
a Toomre diagram is shown in Figure 1. The majority of the M dwarfs show thin disk kinematics
except for a couple of stars showing marginally thick disk kinematics 
($100 <|{\bf V}_{LSR}| < 120$ kms$^{-1}$). There is no star showing halo kinematics.

\section{The Kolmogorov-Smirnov test for C/O ratios obtained by different investigators}

\subsection{Two-sample Kolmogorov-Smirnov test}

Brief introduction to the Kolmogorov-Smirnov (K-S) test is given below \citep{pre88}.
The two-sample K-S test is applicable to a pair of unbinned distributions that are
functions of a single independent variable. In our case, the independent variable is the C/O ratio of 
a star.
We would like to know whether two sets of data are drawn from the same parent distribution function,
or from different distribution functions. The procedure in statistics is itemized in the following.

\noindent
(i) We first set the null hypothesis that data set 1 and data set 2 are derived from the same 
parent distribution. 

\noindent
(ii) We calculate the cumulative probability distribution $S_{N_1}(x)$ of data set 1 and
    the cumulative probability distribution  $S_{N_2}(x)$ of data set 2.

The cumulative distribution $S_N(x)$ is obtained as follows.
{ If there are $N$ data points located at $x_i, i=1,...,N$, which are
sorted into ascending order, $S_N (x)$ is the function giving the fraction of data points
to the left of $x$. $S_N (x)$ is constant between consecutive $x_i$'s and it jumps
by the constant $1/N$ at each $x_i$.}


\noindent
(iii) { 
Different estimates of cumulative distribution function are given by different data sets.
However, all cumulative distribution functions agree at the smallest 
value of $x$ where they are zero, and at the largest value of $x$ where they are unity.
The smallest and largest values of $x$ might be $\pm \infty$.
Therefore the distributions are distinguished by the behavior between the largest
and smallest values.}

\noindent
(iv) We obtain the statistic, Kolmogorov-Smirnov $D$: It is defined as the maximum value
of the absolute difference between two cumulative distribution functions. 

Thus for comparing two different cumulative distribution functions, $S_{N_1}(x)$ and $S_{N_2}(x)$,
the K-S statistic is

 \begin{equation}
 D = \max_{-\infty < x < +\infty} |S_{N_1}(x) - S_{N_2}(x)|.
 \end{equation}

\noindent 
(v) {  The K-S statistic is useful because we can calculate its distribution in the case of
null hypothesis that data sets are drawn from the same distribution.  
Therefore any nonzero value of $D$ is significant.}
The significance level of observed value $D$ (as a disproof of the null hypothesis)
or the probability that $D$ greater than this observed value is obtained by chance,
is given by the formula,

\begin{equation}
{\rm Pr}\left(\sqrt{\frac{N_1 N_2}{N_1 + N_2}}D > z\right) = 2 \sum_{j=1}^\infty (-1)^{j-1} \exp(-2 j^2 z^2),
\end{equation}

where $N_1$ is the number of data points in $S_{N_1}$ and $N_2$ is that in $S_{N_2}$ and the right hand
side is a monotonically decreasing function of $z$.

\noindent
(vi) For a significance level of 1\% (probability = 0.01),
if 

\begin{equation}
\sqrt{\frac{N_1 N_2}{N_1 + N_2}}D > 1.62,
\end{equation}
    
we disprove the null hypothesis and conclude that the two cumulative distributions are derived from different
parent distributions. 

\noindent
(vii) { On the other hand, failing to disprove the null hypothesis, only shows that the data sets can be
 consistent with a single parent distribution function.}

\subsection{Data sets of interest}

We compare using the K-S test pairs of data sets of
cumulative distributions of C/O ratios.
The five data sets of C/O ratios we consider are the following.

\noindent
(a) M dwarfs of this work (46 data points).

Carbon and oxygen abundances have been determined with respect to hydrogen abundance.
These are not the result of a differential abundance analysis with respect to the Sun.
There is no assumption on the C/O ratio of the Sun.

\noindent
(b) \citet{tak05} (149 data points)

The sample consists of nearby disk F,G, and K dwarfs and subgiants.
The abundances of carbon are determined from CI 5052 and 5380 lines.
Non-LTE effect is taken into account for these permitted lines.
They examine three oxygen indicators, the [OI] 6300 (forbidden line),
OI 6158 and OI 7774 triplet for which non-LTE effect
is taken into account and decide to adopt the oxygen abundances
determined from the OI 7774 triplet. 
Their abundance analysis is differential with respect to the Sun
and they present [C/Fe] and [O/Fe] as their final product, from which
[C/O] is calculated. 
They do not make any assumption on solar abundances of carbon and oxygen,
so C/O ratios in absolute scale are not provided.

\noindent
(c) \citet{nis14} (66 data points)

The sample consists of
F and G stars in the Galactic disk. Carbon abundances
are determined from CI 5052 and 5380 lines and
oxygen abundances are determined from the OI 7774 triplet
for which non-LTE corrections are applied.
They also analyze the [OI] 6300 (forbidden line) and
find that the better result is obtained from the OI 7774 triplet. 
These are a product of differential abundance analysis and they
use a solar C/O ratio of 0.58 \citep{nis13} to obtain C/O ratios 
in absolute scale.


\noindent
(d) \citet{del10}  (331 data points)

The sample consists of
F, G, and K stars from HARPS GTO sample \citep{may03}.
Carbon abundances are determined from
CI 5052 and 5380 lines and oxygen abundances are determined from
the  [OI] 6300 forbidden line. The abundance analysis is differential with respect to
the Sun. They adopt a high solar C/O ratio of 0.66 \citep{and89,nis02} to obtain
C/O ratios in absolute scale. 

\noindent
(e) \citet{pet11} (446 data points) 

The sample consists of F and G stars from the SPOCS catalog \citep{val05}
and the N2K sample \citep{fis05}. 
Carbon abundances are determined from
CI 6587 line and oxygen abundances are from [OI] 6300 line. 
They adopt a solar C/O ratio of 0.63 \citep{sco09,caf10} to obtain C/O ratios in absolute scale.

\subsection{Comparison of M dwarf distribution with others}

Since the C/O ratios of M dwarfs are given in absolute scale, the data set to be compared
must also be given in absolute scale. Since no solar abundance ratio is given in the work of \citet{tak05},
it is excluded in this comparison. In the following, the results of comparison between pairs are described.
The cumulative distributions are given graphically in Figure 2, and the parameters of the K-S test
and corresponding panels in Figure 2 are summarized in Table 2.

\subsubsection{
(a) M dwarfs and (c) \citet{nis14} }

The K-S statistic $D = 0.177$ and
the probability that the two data sets are drawn from the same
cumulative distributions is $3.30 \times 10^{-1}$.
This probability does not disprove the null
hypothesis. It shows that the data sets can be consistent with being drawn from
a single cumulative distribution function.

\subsubsection{(a) M dwarfs and (d) \citet{del10} }

The K-S statistic $D = 0.528$ and
the probability that the two data sets are drawn from the same
cumulative distributions is $1.29 \times 10^{-10}$.
This probability is significant enough to disprove the null
hypothesis. It shows that the two data sets are drawn from different cumulative distribution functions. 
The C/O ratios of \citet{del10} are apparently higher than those of M dwarfs.

\subsubsection{(a) M dwarfs and (e) \citet{pet11} }

The K-S statistic $D = 0.427$ and
the probability that the two data sets are drawn from the same
cumulative distributions is $2.60 \times 10^{-7}$.
This probability is significant enough to disprove the null
hypothesis. It shows that the two data sets are drawn from different cumulative distribution functions. 
The C/O ratios of \citet{pet11} are apparently higher than those of M dwarfs.

\subsection{Comparison among the cumulative distributions derived from differential abundance analyses}

{ We consider that both different multiplication factors
(adopted C/O ratios of the Sun) and scatters (dispersions in C/O values)
may be the sources of discrepancies among the C/O distributions of different works in linear scale.
To isolate the effect of scatter,
we compare linear C/O ratios with respect to the Sun, $10^{[{\rm C/O}]}$, calculated from
logarithmic abundances, [C/O] (=[C/H]$-$[O/H]) obtained by differential abundance analyses.}
Since the M dwarf distribution is the result of absolute abundance analysis, it is excluded from
this comparison. 
The cumulative distributions are given graphically in Figures 3 and 4,
and the parameters of the K-S test and corresponding panels in Figures 3 and 4 are summarized in Table 3


\subsubsection{
(b) \citet{tak05} and (c) \citet{nis14} }

The K-S statistic $D = 0.121$ and
the probability that the two data sets are drawn from the same
cumulative distributions is $4.87 \times 10^{-1}$.
This probability is not significant enough to disprove the null
hypothesis. It shows that the data sets can be consistent with being drawn from
a single cumulative distribution function. 

\subsubsection{
(b) \citet{tak05} and (d) \citet{del10} }

The K-S statistic $D = 0.163$ and
the probability that the two data sets are drawn from the same
cumulative distributions is $7.26 \times 10^{-3}$.
This probability is significant enough to disprove the null
hypothesis. It shows that the data sets are drawn from
different cumulative distribution functions.
 
\subsubsection{
(b) \citet{tak05} and (e) \citet{pet11} }

The K-S statistic $D = 0.204$ and
the probability that the two data sets are drawn from the same
cumulative distributions is $1.40 \times 10^{-4}$.
This probability is significant enough to disprove the null
hypothesis. It shows that the data sets are drawn from
different cumulative distribution functions.

\subsubsection{
(c) \citet{nis14} and (d) \citet{del10} }

The K-S statistic $D = 0.234$ and
the probability that the two data sets are drawn from the same
cumulative distributions is $3.80 \times 10^{-3}$.
This probability is significant enough to disprove the null
hypothesis. It shows that the data sets are drawn from
different cumulative distribution functions.

\subsubsection{
(c) \citet{nis14} and (e) \citet{pet11} }

The K-S statistic $D = 0.288$ and
the probability that the two data sets are drawn from the same
cumulative distributions is $1.02 \times 10^{-4}$.
This probability is significant enough to disprove the null
hypothesis. It shows that the data sets are drawn from
different cumulative distribution functions.

\subsubsection{
(d) \citet{del10} and (e) \citet{pet11} }

{
The K-S statistic $D = 0.114$ and
the probability that the two data sets are drawn from the same
cumulative distributions is $1.27 \times 10^{-2}$.
This probability does not disprove the null
hypothesis. It shows that the data sets can be
consistent with being drawn from a single
cumulative distribution.
}


\section{Discussion}




\subsection{C/O ratio of the Sun}

We have seen in the \S3.3 that the distribution of the C/O ratios in M dwarfs can be consistent with
that of \citet{nis14} who use the solar C/O ratio of 0.58,
while it is different from those of \citet{del10} and \citet{pet11}. 
The solar C/O ratio of 0.58 is close to 0.55 recommended by
\citet{asp09}. If the C/O ratio of 0.55 is adopted for \citet{tak05}
and \citet{nis14}, the K-S test with the distribution of M dwarfs gives the probabilities of
both $3.30 \times 10^{-1}$, while $D$ = 0.156 and 0.177 respectively for \citet{tak05} and
\citet{nis14}.
This implies that both cumulative distributions of \citet{tak05} and \citet{nis14} 
can be drawn from the same cumulative distribution as that of M dwarfs for this solar C/O ratio.

For the originally adopted values of solar C/O ratios,
the K-S test of cumulative distributions  
between \citet{nis14} and \citet{del10} gives $D = 0.441$ and the probability = 4.73$\times 10^{-10}$.
For the same K-S test between \citet{nis14} and \citet{pet11} gives $D = 0.338$ and the probability
= 3.32$\times 10^{-8}$. 
The discrepancies between \citet{nis14} and \citet{del10} and between \citet{nis14} and \citet{pet11}
decrease (probabilities increase) for the cases of differential abundance analyses discussed in \S3.4.4 and
\S3.4.5.

If the results for M dwarfs, \citet{tak05} and \citet{nis14} are correct,
the C/O ratios of 0.66 and 0.63 adopted respectively by \citet{del10} and \citet{pet11}
are probably too high.

\subsection{Source of scatter in the C/O ratios of \citet{del10} and \citet{pet11} }

Since the discrepancies of cumulative distributions between 
\citet{del10}, \citet{pet11} and \citet{tak05}, \citet{nis14},
remain in the comparison of differential analyses in \S3.4, the differences in the 
adopted solar C/O ratios alone do not explain these discrepancies.

As discussed by \citet{tak05} and \citet{nis14}, the oxygen abundances derived from the [OI] 6300 forbidden line,
which is blended with Ni line, are most problematic. They compare the results from [OI] 6300 and 
OI 7774 triplet with non-LTE correction and conclude that 
the scatter of oxygen abundances derived from [OI] 6300 is greater than that from OI 7774.  
\citet{del10} and \citet{pet11} both derive oxygen abundances from the [OI] 6300 forbidden line
and this appears to be the major cause of the large scatter in C/O ratios.

{ To visualize the scatter of C/O ratios, we plot [C/O] vs. [Fe/H] diagrams for 
\citet{tak05} and \citet{nis14} in Figure 5, and \citet{del10} and \citet{pet11} in Figure 6.
Strictly speaking, \citet{pet11} give [M/H] instead of [Fe/H], but we use [M/H] as the substitute for
[Fe/H].
These diagrams exhibit the weak trend that C/O ratios are higher for higher [Fe/H],
but also show the much greater scatter in Figure 6 than in Figure 5.}  

{ We also note that the different distributions of [Fe/H] and T$_{\rm eff}$ in the samples
can contribute to the discrepancies in scatter. For example, the stars in \citet{nis14} are
hotter on average than those in \citet{del10}, while those in \citet{pet11} are more metal-rich
on average.}



\subsection{Indirect analysis of C/O ratios in M dwarfs}

\citet{gai15} and \citet{giz16} analyze the C/O ratios in M dwarfs, indirectly from the behavior
of TiO and CaH indices. 
In solar-type stars, the absorption lines of carbon and oxygen are so weak
that the C/O ratio has little effect on the low resolution spectrum.
On the other hand, M dwarf spectra are dominated by simple molecules such as
TiO, VO, and CaOH, which are sensitive to the oxygen abundance available left
after the formation of CO molecules and thus to the C/O ratios.
Both \citet{gai15} and \citet{giz16} use the PHOENIX models \citep{all11,hus13}
to generate model M dwarf spectra for solar composition and for high C/O ratios,
from which qualitative behavior of high C/O ratio stars in the TiO vs. CaH index diagram
is derived. They find that carbon-rich stars can be identified by relatively weak
TiO bands for a given strength of CaH. However, metal poor subdwarfs (sdMs) 
exhibit similar behavior { to carbon-rich stars.}

The work by \citet{gai15} is based on the expectation that
{ at C/O $\ge$ 1, TiO bands are not present, while C$_2$, CN, and CH bands should appear.
He analyzes the spectroscopic catalog of nearby M dwarfs, CONCH-SHELL \citep{gai14}.
He finds that all carbon-rich candidates are either metal poor stars, or have systematic errors.
He estimates that M dwarfs with C/O $\sim$ 1
 constitute less than $1.2 \times 10^{-3}$
with 95\% confidence. He also examines M dwarf spectra in Data Release 7 of SDSS \citep{wes11}
and sets an upper limit of $6 \times 10^{-4}$ at 99\% confidence.}

\citet{giz16} investigate the frequency of M dwarfs with C/O = 0.9 in the solar neighborhood.
{ They first analyze the sample from the complete spectroscopic survey of M dwarfs in the Northern
hemisphere \citep{lep13}, and find that only 2\% of this sample could be 
M dwarfs with high C/O ratios, but many, if not all, are stars with low metallicity. 
Second they analyze the sample from
the PMSU survey \citep{rei95,haw96}. In this sample, 
M subdwarfs (sdM type) are 
less than 1\% and this fraction is 
consistent with the expected numbers of metal-poor (thick disk/halo) population
estimated from kinematics.}
Third they also analyze the SDSS sample \citep{wes11} and find a similar result.
They conclude that less than 1\%
of nearby M dwarfs have $0.8<$C/O$<1$. 


From our analysis of C/O ratios in M dwarfs and solar-type stars, 
the result by \citet{gai15} that M dwarfs with C/O $>$ 1 are very rare appears to be qualitatively
supported. Similarly the result by \citet{giz16} that C/O $\sim$ 0.9 M dwarfs are less than
1\% appears to be supported too. 
However, the methods of analyses are so different that the comparison of the results remains 
only qualitative.


\section{Conclusion}

Recently carbon and oxygen abundances of 46 M dwarfs have been obtained respectively
from CO and H$_2$O molecular lines \citep{tsu14,tsu15,tsu16}. This is not a differential
abundance analysis with respect to the Sun. We present C/O ratios and kinematics of these
M dwarfs. The distribution of C/O ratios in M dwarfs is compared with those in solar type stars
obtained by \citet{nis14}, \citet{del10}, and \citet{pet11} using the K-S test.
The distribution C/O ratios in M dwarfs and that by \citet{nis14} are consistent with
being drawn from a same distribution, while the distributions by \citet{del10} and \citet{pet11}
are not drawn from the same distribution as that in M dwarfs.
High solar C/O ratios adopted by \citet{del10} and \citet{pet11} partly explain these results.

Since carbon and oxygen abundances of solar type stars have been obtained by
differential analyses with respect to the Sun, pairs of C/O ratio distributions with respect
to the solar ratio, are compared among those by \citet{tak05}, \citet{nis14}, \citet{del10}
and \citet{pet11}. { The distributions by \citet{tak05} and \citet{nis14} are consistent
and those by \citet{del10} and \citet{pet11} are barely consistent,
while other pairs are mutually inconsistent.}
Larger scatters in C/O distributions by \citet{del10} and \citet{pet11} are explained by
the use of [OI] 6300 forbidden line as the abundance indicator of oxygen.




\acknowledgments

We thank T. Tsuji for his effort on deriving carbon and oxygen abundances in M dwarfs.
We also thank Y. Takeda for helpful discussions and the anonymous referee for illuminating comments.
This research has made use of the SIMBAD database, operated at CDS, Strasbourg, France.

\clearpage



\begin{figure}
\epsscale{1.0}
\plotone{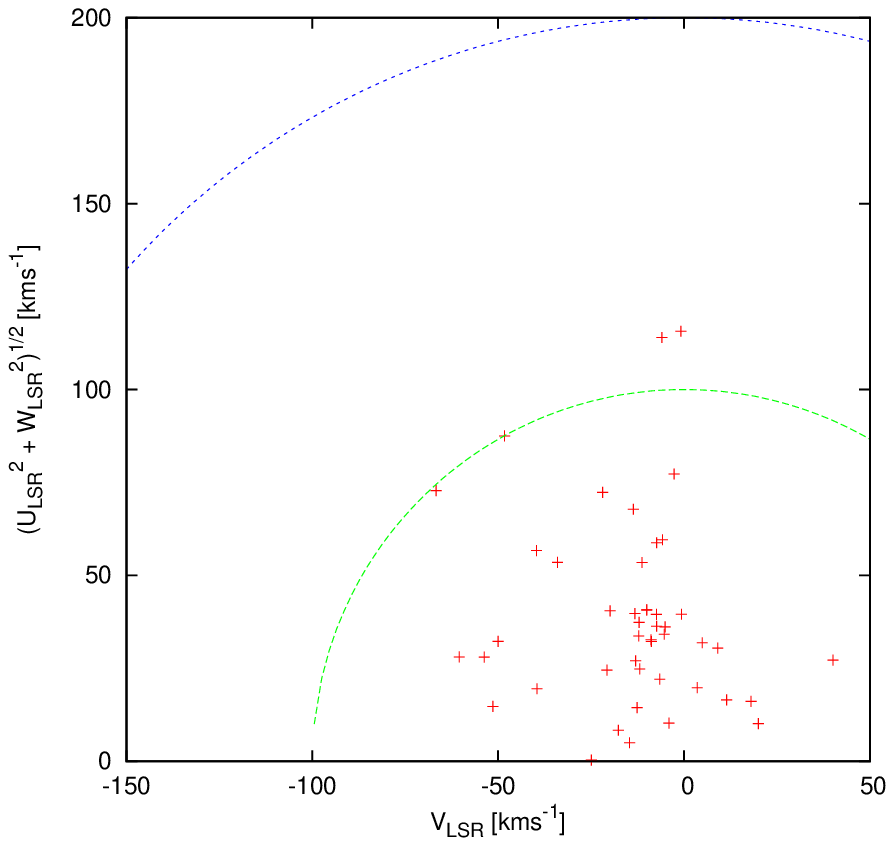}
\caption{Toomre diagram for the sample M dwarfs. The green and blue lines are for
${\bf |V|}_{LSR}$ = 100 and 200 kms$^{-1}$ respectively. Most of the stars have
thin disk kinematics, while two stars possibly have thick disk kinematics (${\bf |V|}_{LSR} > 100$
kms$^{-1}$). There is no star with halo kinematics.  \label{fig1}}
\end{figure}

\clearpage


\begin{figure}
\plotone{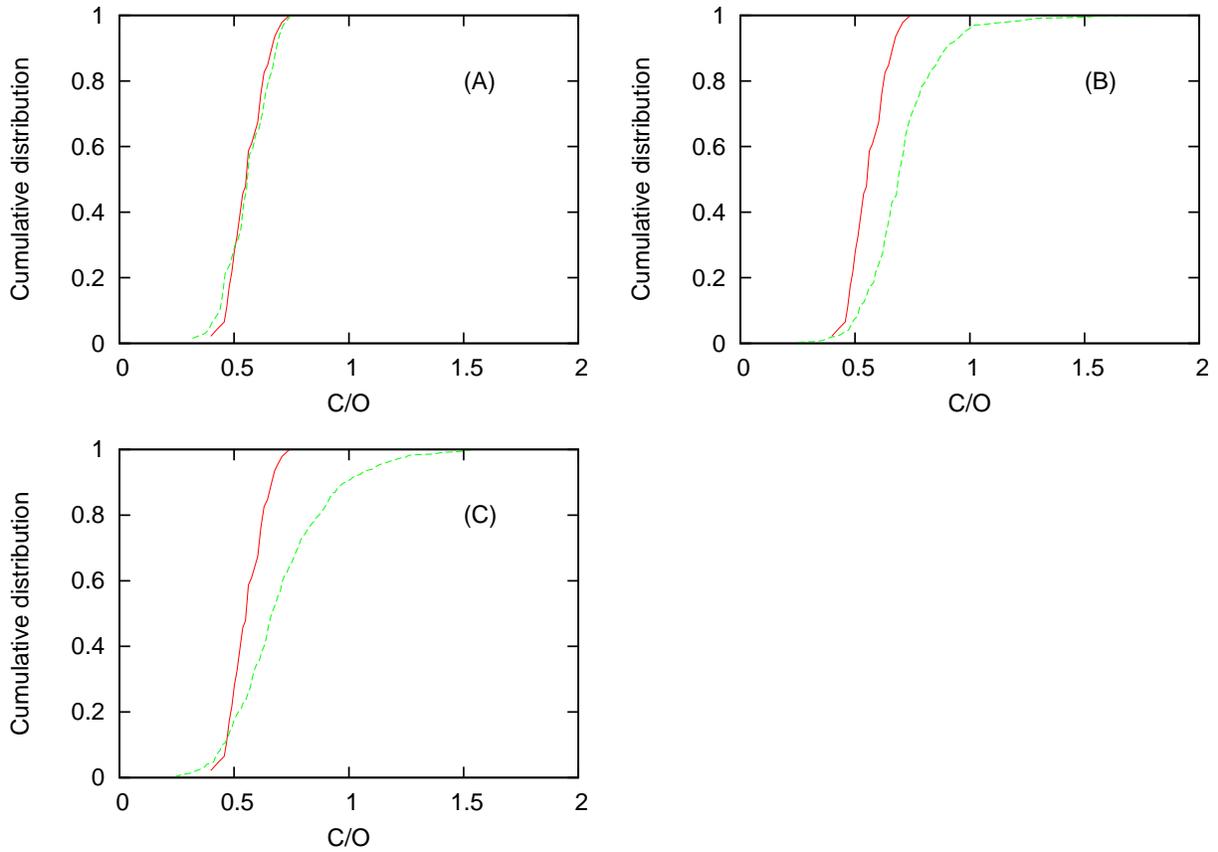}
\caption{Comparison of two cumulative distributions. 
{ (A) M dwarfs (red) and \citet{nis14}. (B) M dwarfs (red) and \citet{del10} (green).
(C) M dwarfs (red) and \citet{pet11}}.
The probability that each pair are the same is given in
Table 2. \label{fig2}}
\end{figure}

\clearpage

\begin{figure}
\plotone{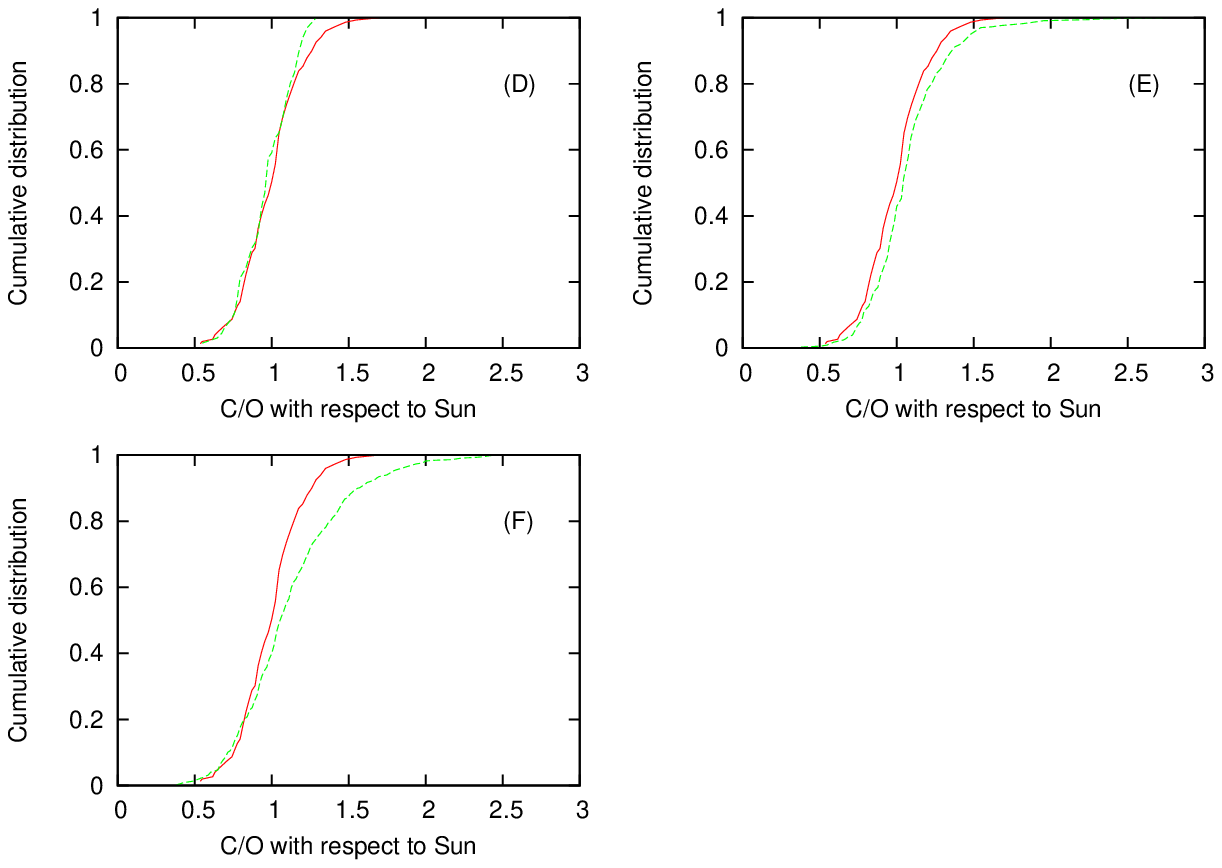}
\caption{Comparison of two cumulative distributions. 
{ (D) \citet{tak05} (red) and \citet{nis14} (green). (E) \citet{tak05} (red) and \citet{del10} (green).
(F) \citet{tak05} (red) and \citet{pet11} (green).}  
The probability that each pair are the same is given in
Table 3. \label{fig3}}
\end{figure}

\clearpage

\begin{figure}
\plotone{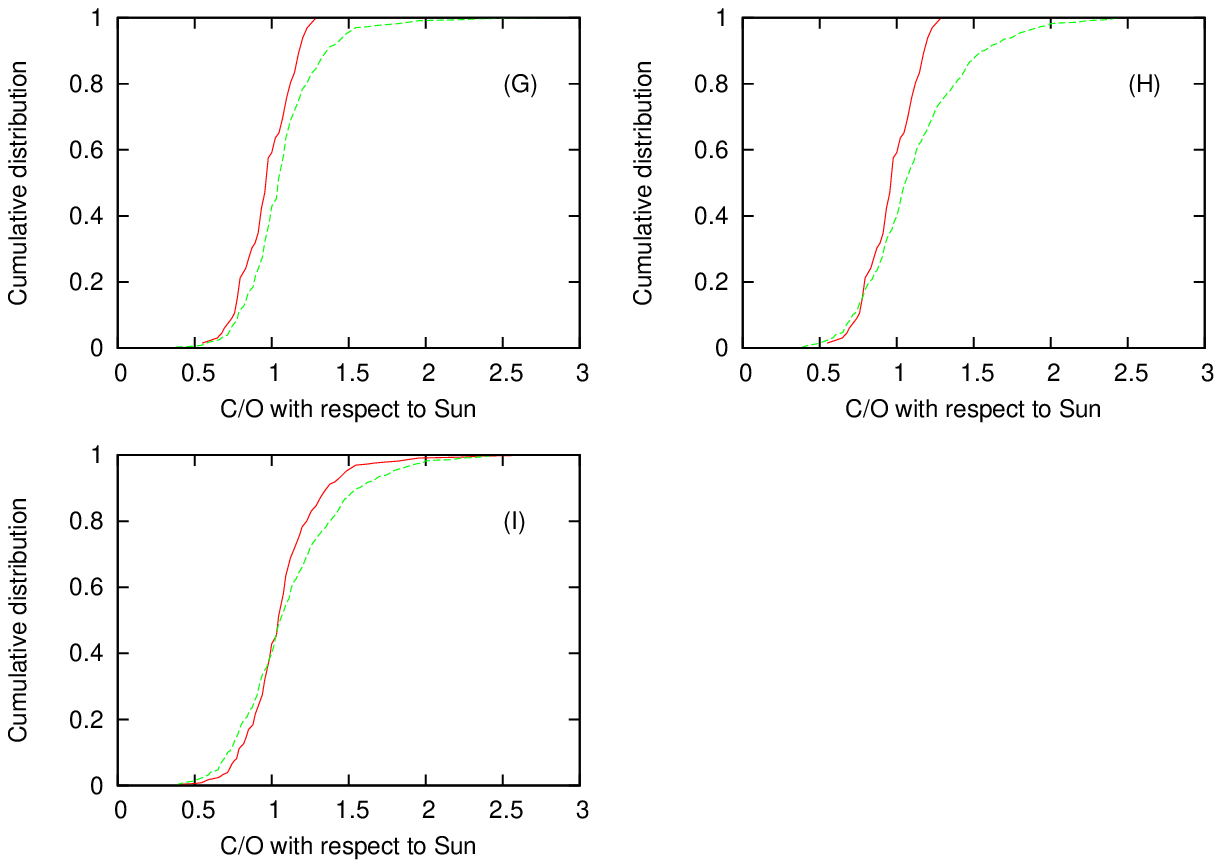}
\caption{Comparison of two cumulative distributions. 
{ (G) \citet{nis14} (red) and \citet{del10} (green). (H) \citet{nis14} (red) and \citet{pet11} (green).
(I) \citet{del10} (red) and \citet{pet11} (green).}  
The probability that each pair are the same is given in
Table 3. \label{fig4}}
\end{figure}

\begin{figure}
\plotone{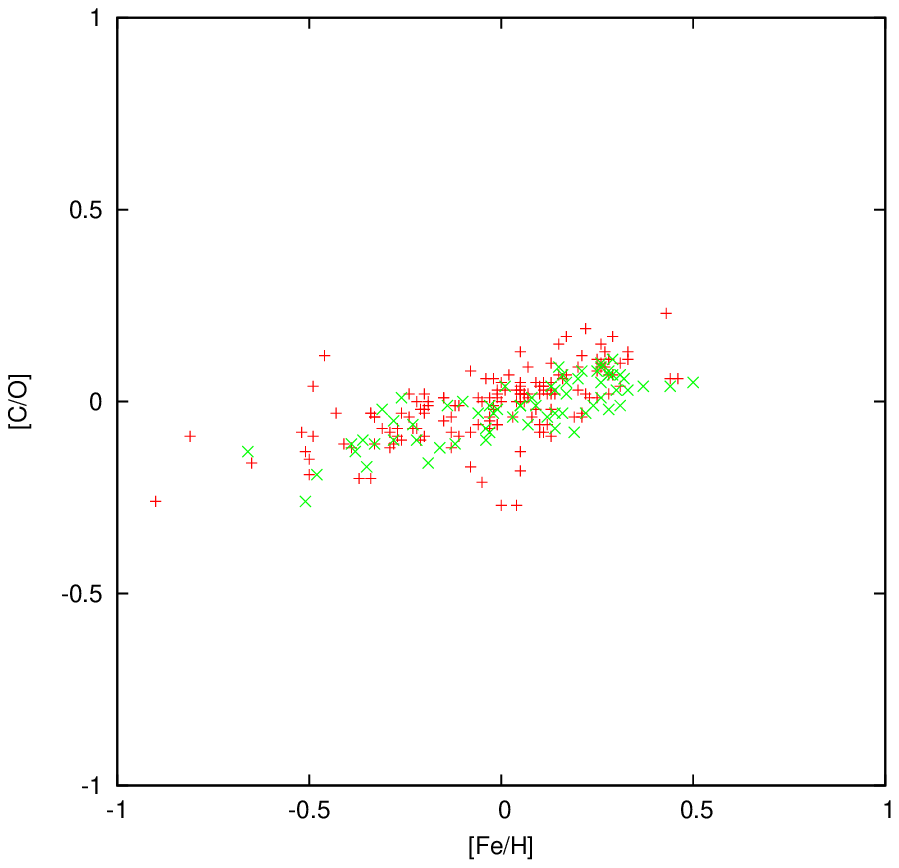}
\caption{[C/O] vs. [Fe/H] for \citet{tak05} (red) and \citet{nis14} \label{fig5} (green).}
\end{figure}

\begin{figure}
\plotone{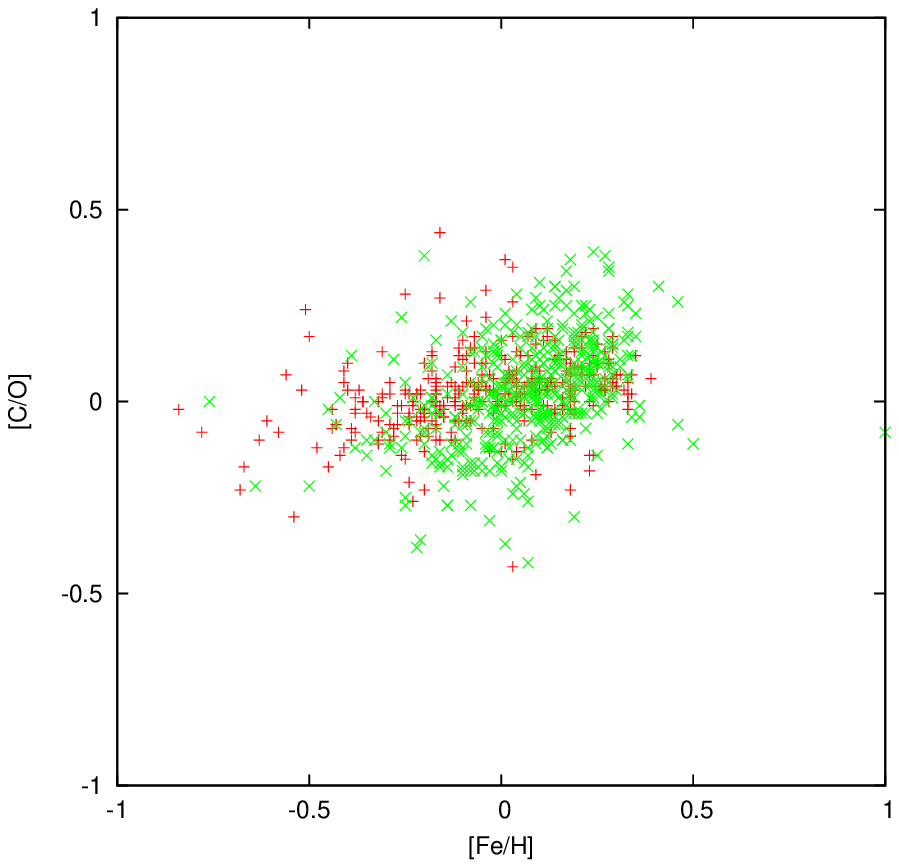}
\caption{[C/O] vs. [Fe/H] for \citet{del10} (red) and \citet{pet11} (green) \label{fig6}.
For \citet{pet11}, [M/H] is used as the substitute for [Fe/H].}
\end{figure}









\clearpage

\begin{deluxetable}{cccccccc}
\tabletypesize{\scriptsize}
\tablecaption{Carbon and oxygen abundances and kinematics\label{tbl-1}}
\tablewidth{0pt}
\tablehead{
\colhead{Object} & \colhead{$\log A_C$} & \colhead{$\log A_O$} & \colhead{C/O}  &
\colhead{$U_{LSR}$} & \colhead{$V_{LSR}$} & \colhead{$W_{LSR}$} &
\colhead{$|{\bf V}_{LSR}|$} \\
\colhead{} & \colhead{} & \colhead{} & \colhead{}  & 
\colhead{kms$^{-1}$} & \colhead{kms$^{-1}$} & \colhead{kms$^{-1}$} & \colhead{kms$^{-1}$} 
}
\startdata
       GJ15A &  -3.60$\pm$0.11 &  -3.31$\pm$0.01 & 0.51  &     -39.3 &    -7.5 &     4.3 &    40.2 \\ 
      GJ105B &  -3.47$\pm$0.06 &  -3.14$\pm$0.04 & 0.47  &     -69.2 &    -2.7 &    34.3 &    77.3 \\ 
      GJ166C &  -3.37$\pm$0.14 &  -3.03$\pm$0.06 & 0.46  &     108.8 &    -6.0 &   -34.0 &   114.1 \\ 
       GJ176 &  -3.35$\pm$0.07 &  -3.15$\pm$0.02 & 0.63  &     -12.8 &   -51.4 &    -7.3 &    53.5 \\ 
       GJ179 &  -3.43$\pm$0.11 &  -3.18$\pm$0.04 & 0.56  &      23.5 &   -11.9 &     8.0 &    27.5 \\ 
       GJ205 &  -3.12$\pm$0.08 &  -2.97$\pm$0.05 & 0.71  &      32.1 &   -50.0 &    -3.2 &    59.5 \\ 
       GJ212 &  -3.30$\pm$0.11 &  -3.10$\pm$0.03 & 0.63  &      -4.6 &   -17.6 &    -7.0 &    19.5 \\ 
       GJ229 &  -3.27$\pm$0.07 &  -3.10$\pm$0.02 & 0.68  &      21.6 &    -6.6 &    -4.8 &    23.1 \\ 
    GJ231.1B &  -3.57$\pm$0.05 &  -3.24$\pm$0.05 & 0.47  &     -10.0 &    20.0 &    -1.6 &    22.4 \\ 
      GJ250B &  -3.41$\pm$0.10 &  -3.19$\pm$0.03 & 0.60  &       9.5 &    18.0 &   -13.0 &    24.1 \\ 
       GJ273 &  -3.40$\pm$0.11 &  -3.13$\pm$0.04 & 0.54  &      26.2 &   -60.5 &    -9.9 &    66.7 \\ 
      GJ324B &  -3.36$\pm$0.13 &  -3.17$\pm$0.04 & 0.65  &     -27.0 &   -13.1 &     0.1 &    30.0 \\ 
      GJ338A &  -3.59$\pm$0.04 &  -3.32$\pm$0.03 & 0.54  &     -29.4 &    -8.8 &   -13.2 &    33.4 \\ 
      GJ338B &  -3.58$\pm$0.04 &  -3.31$\pm$0.06 & 0.54  &     -34.1 &   -12.1 &   -15.2 &    39.3 \\ 
       GJ406 &  -3.61$\pm$0.10 &  -3.40$\pm$0.03 & 0.62  &     -16.1 &   -39.6 &   -11.0 &    44.1 \\ 
       GJ411 &  -3.67$\pm$0.06 &  -3.37$\pm$0.03 & 0.50  &      55.7 &   -48.3 &   -67.5 &   100.0 \\ 
      GJ412A &  -3.85$\pm$0.04 &  -3.53$\pm$0.03 & 0.48  &    -113.0 &    -0.9 &    24.8 &   115.7 \\ 
       GJ436 &  -3.63$\pm$0.06 &  -3.42$\pm$0.02 & 0.62  &      62.5 &   -13.7 &    26.3 &    69.2 \\ 
       GJ526 &  -3.55$\pm$0.04 &  -3.28$\pm$0.03 & 0.54  &      59.1 &    -5.8 &     7.1 &    59.8 \\ 
       GJ581 &  -3.56$\pm$0.05 &  -3.26$\pm$0.03 & 0.50  &     -14.7 &   -20.7 &    19.6 &    32.1 \\ 
      GJ611B &  -3.76$\pm$0.03 &  -3.36$\pm$0.06 & 0.40  &     -25.9 &   -53.7 &   -10.6 &    60.6 \\ 
       GJ649 &  -3.54$\pm$0.04 &  -3.34$\pm$0.03 & 0.63  &      31.6 &    -8.9 &     8.3 &    33.9 \\ 
       GJ686 &  -3.50$\pm$0.04 &  -3.23$\pm$0.03 & 0.54  &     -23.6 &    40.1 &   -13.5 &    48.4 \\ 
       GJ687 &  -3.43$\pm$0.09 &  -3.22$\pm$0.03 & 0.62  &      40.5 &   -19.9 &     0.2 &    45.1 \\ 
      GJ725A &  -3.58$\pm$0.09 &  -3.27$\pm$0.05 & 0.49  &     -14.7 &    -7.4 &    33.2 &    37.1 \\ 
      GJ725B &  -3.61$\pm$0.08 &  -3.29$\pm$0.06 & 0.48  &     -13.8 &    -5.1 &    33.4 &    36.5 \\ 
    GJ768.1C &  -3.50$\pm$0.08 &  -3.20$\pm$0.04 & 0.50  &       8.2 &     3.6 &   -18.0 &    20.1 \\ 
      GJ777B &  -3.24$\pm$0.16 &  -3.02$\pm$0.05 & 0.60  &      -2.5 &   -39.7 &   -56.6 &    69.2 \\ 
    GJ783.2B &  -3.41$\pm$0.10 &  -3.14$\pm$0.05 & 0.54  &     -20.6 &   -21.9 &    69.3 &    75.6 \\ 
   GJ797B-NE &  -3.54$\pm$0.09 &  -3.28$\pm$0.03 & 0.55  &     -32.8 &   -10.1 &    24.1 &    42.0 \\ 
   GJ797B-SW &  -3.51$\pm$0.09 &  -3.26$\pm$0.03 & 0.56  &     -32.8 &   -10.1 &    24.1 &    42.0 \\ 
       GJ809 &  -3.55$\pm$0.04 &  -3.37$\pm$0.03 & 0.66  &      31.9 &    -5.3 &   -12.2 &    34.6 \\ 
       GJ849 &  -3.27$\pm$0.09 &  -3.09$\pm$0.02 & 0.66  &     -32.6 &   -12.2 &    -8.4 &    35.9 \\ 
       GJ876 &  -3.36$\pm$0.13 &  -3.14$\pm$0.04 & 0.60  &      -2.4 &   -14.7 &    -4.4 &    15.5 \\ 
       GJ880 &  -3.35$\pm$0.07 &  -3.18$\pm$0.03 & 0.68  &      42.7 &   -11.3 &    32.1 &    54.6 \\ 
     GJ3348B &  -3.40$\pm$0.11 &  -3.08$\pm$0.05 & 0.48  &       0.3 &   -25.0 &     0.1 &    25.0 \\ 
    HIP57050 &  -3.26$\pm$0.12 &  -3.05$\pm$0.04 & 0.62  &     -12.0 &   -12.6 &    -8.0 &    19.2 \\ 
    HIP79431 &  -3.24$\pm$0.12 &  -3.11$\pm$0.01 & 0.74  &       9.1 &    -4.1 &    -4.7 &    11.0 \\ 
      GJ54.1 &  -3.38$\pm$0.09 &  -3.13$\pm$0.06 & 0.56  &      -8.3 &    -0.7 &    38.7 &    39.6 \\ 
      GJ752B &  -3.55$\pm$0.07 &  -3.31$\pm$0.04 & 0.58  &      58.8 &    -7.4 &     1.7 &    59.2 \\ 
       GJ873 &  -3.44$\pm$0.09 &  -3.13$\pm$0.06 & 0.49  &      30.0 &     9.1 &     5.2 &    31.8 \\ 
      GJ1002 &  -3.42$\pm$0.09 &  -3.17$\pm$0.05 & 0.56  &      46.5 &   -34.0 &    26.5 &    63.4 \\ 
     GJ1245B &  -3.44$\pm$0.10 &  -3.15$\pm$0.07 & 0.51  &      15.9 &    11.4 &    -4.5 &    20.1 \\ 
     GAT1370 &  -3.78$\pm$0.06 &  -3.41$\pm$0.06 & 0.43  &     -55.4 &   -66.7 &   -47.2 &    98.7 \\ 
    LP412-31 &  -3.36$\pm$0.07 &  -3.20$\pm$0.03 & 0.69  &     -36.9 &   -13.2 &   -14.8 &    41.9 \\ 
   2M1835+32 &  -3.57$\pm$0.20 &  -3.32$\pm$0.03 & 0.56  &      31.7 &     4.9 &     3.3 &    32.3 
\enddata
\end{deluxetable}


\clearpage

\begin{table}
\begin{center}
\caption{The Kolmogorov-Smirnov test between M dwarfs and others\label{tbl-2}}
\begin{tabular}{ccccc}
\tableline\tableline
Distribution 1 & Distribution 2 & $D$ &  Probability\tablenotemark{a} & Panel \tablenotemark{b} \\
\tableline
  M dwarfs  &   Nissen et al.      & 0.177 &  3.30$\times10^{-1}$          & A \\
  M dwarfs  &  Delgado Mena et al. & 0.528 &  $1.29 \times 10^{-10}$ & B \\
  M dwarfs  &   Petigura \& Marcy   & 0.427 &  2.60$\times10^{-7}$       & C \\
\tableline
\end{tabular}
\tablenotetext{a}{Probability that the two distributions are the same. If this value is less than
0.01, the two distribution are regarded as different.}
\tablenotetext{b}{Panel in Figure 2 In each panel, distribution 1 is in red and distribution 2 
is in green.}
\end{center}
\end{table}

\clearpage

\begin{table}
\begin{center}
\caption{The Kolmogorov-Smirnov test among differential analyses\label{tbl-3}}
\begin{tabular}{ccccc}
\tableline\tableline
Distribution 1 & Distribution 2  & $D$ &  Probability\tablenotemark{a} & Panel \tablenotemark{b} \\
\tableline
   Takeda \& Honda  &   Nissen et al. & 0.121 & 4.87$\times10^{-1}$          & D \\
   Takeda \& Honda  &   Delgado Mena et al. & 0.163 & $7.26 \times 10^{-3}$  & E \\
  Takeda \& Honda  &    Petigura \& Marcy   & 0.204 & 1.40$\times10^{-4}$    & F \\
   Nissen et al.    &  Delgado Mena et al. & 0.234 & $3.80 \times 10^{-3}$  & G \\
   Nissen et al.    &   Petigura \& Marcy   & 0.288 &  $1.02 \times 10^{-4}$  & H \\
  Delgado Mena et al. &   Petigura \& Marcy & 0.114 & $1.27 \times 10^{-2}$  & I \\
\tableline
\end{tabular}
\tablenotetext{a}{Probability that the two distributions are the same. If this value is less than
0.01, the two distribution are regarded as different.}
\tablenotetext{b}{Panel in Figures 3 and 4. In each panel, distribution 1 is in red and distribution 2 
is in green.}
\end{center}
\end{table}

\end{document}